\documentclass[]{aa}  

\usepackage{natbib}
\bibpunct{(}{)}{;}{a}{}{,}
\usepackage{graphicx}
\usepackage[export]{adjustbox}
\usepackage{revsymb}    
\usepackage{amsmath}    
\usepackage[usenames]{color}    
\usepackage{txfonts}
\usepackage{amssymb}
\usepackage{color}
\usepackage{microtype}
\usepackage{geometry} 
\usepackage[parfill]{parskip} 
\usepackage{amssymb}
\usepackage{slantsc}
\usepackage{array}
\usepackage{url}
\usepackage{amsfonts}
\usepackage[colorlinks,linkcolor=blue]{hyperref}
\hypersetup{                    
    colorlinks=true,            
    breaklinks=true,            
    urlcolor= blue,             
    linkcolor= black,           
    citecolor= blue             
    }
\usepackage[nameinlink,capitalise]{cleveref}

\usepackage{sidecap}
\usepackage{xcolor}
\usepackage{nicefrac}
\usepackage{subfigure}
\usepackage{epsfig}
\colorlet{rouge}{red!70!darkgray}

\let\orgautoref\autoref
\providecommand{\Autoref}
        {\def\equationautorefname{Equation}%
         \def\figureautorefname{Figure}%
         \def\subfigureautorefname{Figure}%
         \def\sectionautorefname{Section}%
         \def\subsectionautorefname{Section}%
         \def\subsubsectionautorefname{Section}%
         \def\Itemautorefname{Item}%
         \def\tableautorefname{Table}%
         \orgautoref}

\renewcommand{\autoref}
        {\def\equationautorefname{Eq.}%
         \def\figureautorefname{Fig.}%
         \def\subfigureautorefname{Fig.}%
         \def\sectionautorefname{Sect.}%
         \def\subsectionautorefname{Sect.}%
         \def\subsubsectionautorefname{Sect.}%
         \def\Itemautorefname{item}%
         \def\tableautorefname{Table}%
         \orgautoref}

\usepackage{threeparttable}

\begin{document}

   \title{Coralie radial-velocity search for companions\\around evolved stars (CASCADES)}
      \subtitle{III. A new Jupiter host-star: in-depth analysis of HD\,29399 using TESS data. \thanks{Based on observations collected with the CORALIE echelle spectrograph on the 1.2-m Euler Swiss telescope at La Silla Observatory, ESO, Chile.}}
\author{C. Pezzotti\inst{1} \and G. Ottoni\inst{1} \and G. Buldgen\inst{1} \and A. Lyttle\inst{2} \and P. Eggenberger\inst{1} \and S. Udry\inst{1} \and D. Ségransan\inst{1} \and M. Mayor\inst{1} \and C. Lovis\inst{1} \and M. Marmier\inst{1} \and A. Miglio\inst{2} \and Y. Elsworth\inst{2} \and G.R. Davies \inst{2} \and W.H. Ball\inst{2}}
\institute{Département d'Astronomie, Université de Genève, Chemin Pegasi 51, 1290 Versoix, Switzerland
\and School of Physics and Astronomy, University of Birmingham, Edgbaston, Birmingham B15 2TT, UK.}

   \date{Received December 7, 2020; accepted ...}

\abstract
{Increasing the number of detected exoplanets is far from anecdotal, especially for long-period planets that require a long duration of observation. More detections imply a better understanding of the statistical properties of exoplanet populations, and detailed modelling of their host stars also enables thorough discussions of star--planet interactions and orbital evolution of planetary systems.}
{In the context of the discovery of a new planetary system, we aim to perform a complete study of HD\,29399 and its companion by means of radial-velocity measurements, seismic characterisation of the host-star, and modelling of the orbital evolution of the system.}
{High-resolution spectra of HD\,29399 were acquired with the CORALIE spectrograph mounted on the 1.2-m Swiss telescope located at La Silla Observatory (Chile) as part of the CASCADES survey. We used the moments of the cross-correlation function  profile as well as the photometric variability of the star as diagnostics to distinguish between stellar and planetary-induced signals. To model the host star we combined forward modelling with global and local minimisation approaches and inversion techniques. We also studied the orbital history of the system under the effects of both dynamical and equilibrium tides.}
{We present the detection of a long-period giant planet. Combining these measurements with photometric observations by TESS, we are able to thoroughly model the host star and study the orbital evolution of the system. We  derive stellar and planetary masses of $\rm 1.17\pm0.10 ~M_{\odot}$ and $1.59 \pm 0.08~ \rm{M_{Jup}}$, respectively, and an age for the system of $6.2$ Gyr. We show that neither dynamical nor equilibrium tides have been able to affect the orbital evolution of the planet. Moreover, no engulfment is predicted for the future evolution of the system.}
{}


\maketitle
\section{Introduction}

Owing to the discovery of more than 4300 exoplanets, the booming field of exoplanetology has seen tremendous success and encountered rapid development. Today, the attention of the field is not only focused on the detection of distant worlds but also on the characterisation of exoplanetary populations \citep{Udry2007,Udry2010,Winn2015}, revelation of their formation and orbital history \citep{Mulders2018,Jontof2019}, and the detailed study of their atmospheric signatures \citep{Seager2010,Kaltenegger2017,Fujii2018}. 

The detection and study of planets around giant stars is particularly interesting, providing information about the architecture and evolution of systems orbiting around stars more evolved than our Sun. The advantage of observing more massive stars at later stages of  evolution, in this case along the red giant branch (RGB), is that they have a decreased surface temperature and slower surface rotation rate compared to what is observed during the pre-main sequence (PMS) and main-sequence (MS) phases. This has the effect of increasing the number of absorption lines and their sharpness (because of rotational broadening) in the spectra, thus enabling the measurement of precise stellar radial velocities (RV) suitable for exoplanet searches. The trend in searching for planets around evolved intermediate-mass stars began with the announcement of a planet orbiting the K2 III giant $\iota$ Draconis \citep{Frink2002}, and led to the discovery of more than 100 systems. However, the correct analysis and interpretation of the radial-velocity variations remains challenging because of the significant and potentially periodic intrinsic variability of red giants, which can mimic planetary signals \citep[see][for more detailed discussion on the different potential origins of this variability]{Walker1989, Hatzes1993, Hatzes1994, Hatzes1999, Frandsen2002, DeRidder2006, Hekker2006a}. In this context, the CORALIE radial-velocity search for companions around evolved stars (Ottoni 2021 submitted, CASCADES) started monitoring a sample of more than 600 G- and K-type giant stars in the southern hemisphere in 2006 (see \autoref{subsubsec:cascades}).

Probing more massive stars is also useful for discussions of the main competing planet formation scenarios. Indeed, stellar mass has a significant impact on the evolution of the protostellar disc \citep{Ribas2015}, and thus its lifetime and ability to form giant planets: a more massive star with a shorter disc lifetime would favour the disc instability scenario, which can form massive planets on very short timescales of a few thousand years \citep[e.g.][]{Helled2014,Raymond2014}.

While the field of exoplanetology was undergoing its rapid evolution, another area of research started acquiring importance, asteroseismology. Rapidly, synergies between exoplanetary studies and the seismic characterisation of stars became a standard example of successful multidisciplinary studies \citep[e.g.][]{JCDKOI2010,Batalha2011,Huber2013,Huber2013Science,Huber2018,Campante2018}. Indeed, the precise characterisation of distant worlds required access to reliable and precise stellar parameters, such as mass, radius, and age. With the development of sophisticated seismic modelling techniques, such achievements become feasible, especially owing to the space-based photometry missions such as CoRoT, \textit{Kepler}, TESS, and, in the future, PLATO. Adding the recent publication of \textit{Gaia} DR2 \citep{Evans2018, Gaia2018}, the achievable precision and accuracy of such detailed studies has reached unprecedented levels for a large number of stars.  


We start in \autoref{sec:obs_stellprop} with a brief description of the CASCADES survey and the method used for determining stellar parameters. We also present the acquisition and analysis of the spectroscopic measurements, and a study of the intrinsic variability of the star HD\,29399 using both spectroscopic and photometric data. In \autoref{sec:kep_analysis} we present the orbital solution of our new planetary companion. \Autoref{SecAstero} details the asteroseismic analysis, from the peak-bagging procedure to obtain individual oscillation frequencies for the host star to the thorough modelling that we carry out, combining forward modelling and local and global minimisation techniques with seismic inversions. Finally, in \autoref{SecOrbital}, we investigate both the role of dynamical tides during the PMS and the role of equilibrium tides on the MS and the RGB using the stellar parameters determined in Section 4.

\section{Observations, stellar properties, and intrinsic variability}\label{sec:obs_stellprop}

\subsection{Observations and stellar parameters}
\subsubsection{The CASCADES survey}\label{subsubsec:cascades}
The CORALIE radial-velocity Search for Companions ArounD Evolved Stars (CASCADES) is a 14-year survey of a volume-limited sample of evolved stars of intermediate mass. The main motivation for this survey was to better understand the formation of planetary systems and their evolution around stars more massive than the Sun by completing existing studies of giant host stars and their companions. Observations began at the end of 2006, and have been conducted since then with the CORALIE spectrograph on the 1.2-m Leonard Euler Swiss telescope located at La Silla Observatory (Chile). 
For a detailed description of the definition of the sample and first results of the survey, see Ottoni et al. (2021, submitted), hereafter Paper I. Complete information on instrumental aspects is given for instance in \citet{Queloz2000}, \citet{Segransan2010}, and Ségransan et al. (2020, submitted).

We collected 28 radial-velocity measurements for HD\,29399 with the CORALIE spectrograph over a time-span of more than 13 years. The obtained spectra have an average signal-to-noise ratio of 85 (at 5'500\,\AA) for typical exposure times of $\sim$180\,s. \autoref{tab:timeseries_hd29399} gives the list of these radial-velocity measurements and uncertainties.

\subsubsection{Additional measurements}
An additional set of measurements was used for the analysis: four unpublished data points from the HARPS spectrograph \citep{Mayor2003} (mounted on the 3.6 m ESO telescope at La Silla Observatory, Chile), and 22 radial-velocity points published by \citet{Wittenmyer2017a} from the UCLES spectrograph \citep{Diego1990} on the AAT  and the CHIRON spectrograph \citep{Tokovinin2013} on the 1.5 m telescope at CTIO. 

\subsubsection{Spectroscopic parameters}
The spectroscopic parameters of the stars in the CASCADES sample are provided in Paper I. Following the method described in \citet{Alves2015}, we derived the effective temperature $T_{\rm{eff}}$, surface gravity $\rm{log\,g,}$ and metallicity $\rm{[Fe/H]}$ using the CORALIE high-resolution spectra. The values obtained were in agreement with those found for the subsample of stars in common with \citet{Alves2015}, and with the values from the \textit{Gaia}-DR2 \citep{Brown2018}. The spectroscopic parameters derived for HD\,29399, along with other stellar parameters obtained from the literature, are presented in \autoref{tab:stellar_params}.

\begin{table}[t]
\centering
\begin{threeparttable}
\caption{Observed and inferred stellar parameters.}
    \begin{tabular}{lllc}        
    \hline
    \hline
    & & ref. & HD\,29399\\
    & & TIC & 38828538\\
    & & GAIA DR2 & {\tiny 4675576135153914368}\\
    \hline
    Sp. Type & & [1] & K1III \\
    $V$ & [mag] & [2] & 5.79 $\pm$ 0.01 \\
    $B-V$ & [mag] & [2] & 1.03 $\pm$ 0.01 \\
    $BC$ & & [3] & -0.317 $\pm$ 0.019 \\
    $\pi$ & [mas] & [4] & 22.62~$\pm$~0.05 \\
    $d$ & [pc] & [5] & 44.2 $\substack{+0.1 \\ -0.1}$ \\
    $M_{V}$ & [mag] & [2,4,5] & 2.56 $\pm$ 0.01 \\
    $Bp-Rp$ & [mag] & [4] & 1.170 $\pm$ 0.003 \\
    $G$ & [mag] & [4] & 2.26 $\pm$ 0.01 \\
    $T_{eff}$ & [K] & [4] & 4803 $\substack{+66 \\ -64}$ \\
    & & [6] & 4845 $\pm$ 52 \\
    $log\,g$ & [cm\,s$^{-2}$] & [6] & 3.25 $\pm$ 0.13 \\
    $[Fe/H]$ & [dex] & [6] & 0.14 $\pm$ 0.03 \\
    $M_{*}$ & [M$_{\odot}$] & [7] & 1.17 $\pm$ 0.10 \\
    $L_{*}$ & [L$_{\odot}$] & [2,3,4] & 10.04 $\pm$ 0.20 \\
    $R_{*}$ & [R$_{\odot}$] & [2,3,4,6] & 4.50 $\pm$ 0.11 \\
    \hline
    Mass$_{opt}$ & [M$_{\odot}$] & [8] & 1.15 $\pm$ 0.04\\
    Radius$_{opt}$ & [R$_{\odot}$] & [8] & 4.47 $\pm$ 0.02\\
    X$_0$ & & [8] & 0.681 $\pm$ 0.01\\
    Z$_0$ & & [8] & 0.0170 $\pm$ 0.001\\
    $\alpha_{\rm{MLT}}$ & [H$_{P}$] & [8] & 2.00 $\pm$ 0.05\\
    Age & [Gy] & [8] & 6.20 $\pm$ 0.5\\
    \hline
    \end{tabular}
\begin{tablenotes}
    \small
    \item {[1]} - HIPPARCOS catalogue \citep{ESA1997}, [2] - TYCHO-2 catalogue \citep{Hog2000}, [3] - \citet{Alonso1999}, [4] - Gaia DR2 \citep{Brown2018}, [5] - \citet{Bailer-Jones2018}, [6] - this paper (see \autoref{sec:obs_stellprop}), [7] - model-independent mass from seismic inversion (see \autoref{SecAstero}), [8] parameter of the optimal stellar model (see \autoref{SecAstero}).
\end{tablenotes}
\label{tab:stellar_params}
\end{threeparttable}
\end{table}

\begin{figure}[t!]
        \centering
        \adjincludegraphics[width=1\columnwidth, trim={0 0 0 0},clip]{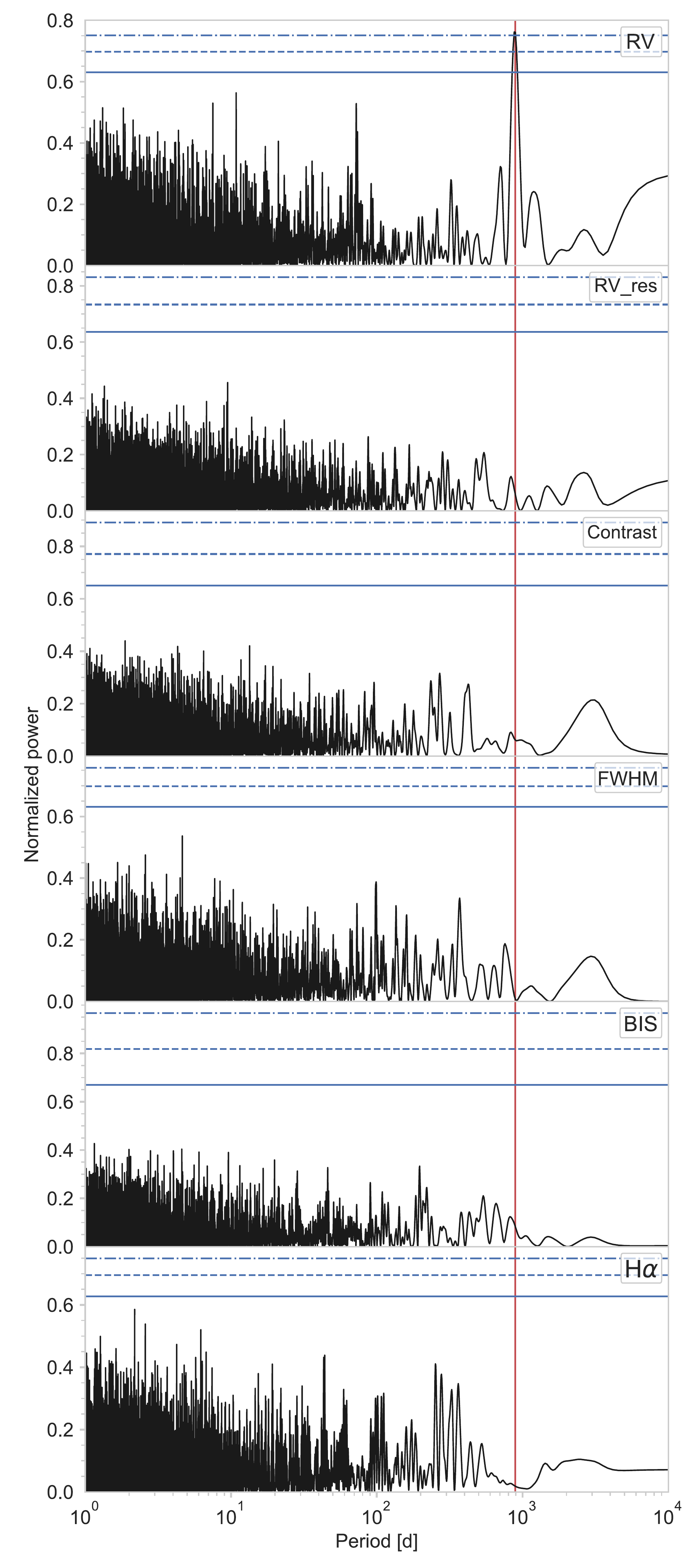}
        \caption{Periodogram of the radial-velocity data (first panel), of the residuals of the radial velocities after subtraction of the fitted periodic signal (second panel), of the contrast (third panel), of the FWHM (fourth panel), of the bisector inverse span (fifth panel), and of the equivalent width of H$\alpha$ activity (sixth panel). The red vertical line represents the fitted period in the radial-velocity at 892.7 days. Horizontal lines, from bottom to top, are the FAP levels at 10\%, 1\%, and 0.1\% respectively.}
        \label{fig:spec_perio}
\end{figure}

\subsection{Stellar analysis of HD~29399}\label{subsec:hd29399_obs}
We first analysed the radial-velocity time series using the radial-velocity module of the DACE web platform, which provides open access to a wide range of  observational and theoretical exoplanet data with the corresponding data visualisation and analysis tools\footnote{\href{https://dace.unige.ch/radialVelocities/?}{https://dace.unige.ch/radialVelocities/?}. 
The formalism of the radial-velocity data analysis implemented in DACE is described in Ségransan et al. (2021, submitted) and is mainly based on algorithms presented in \citet{Diaz2014} and \citet{Delisle2016, Delisle2018}}.

\begin{figure}[t]
        \centering
        \adjincludegraphics[width=.9\columnwidth, trim={0 0 0 {0.025\height}},clip]{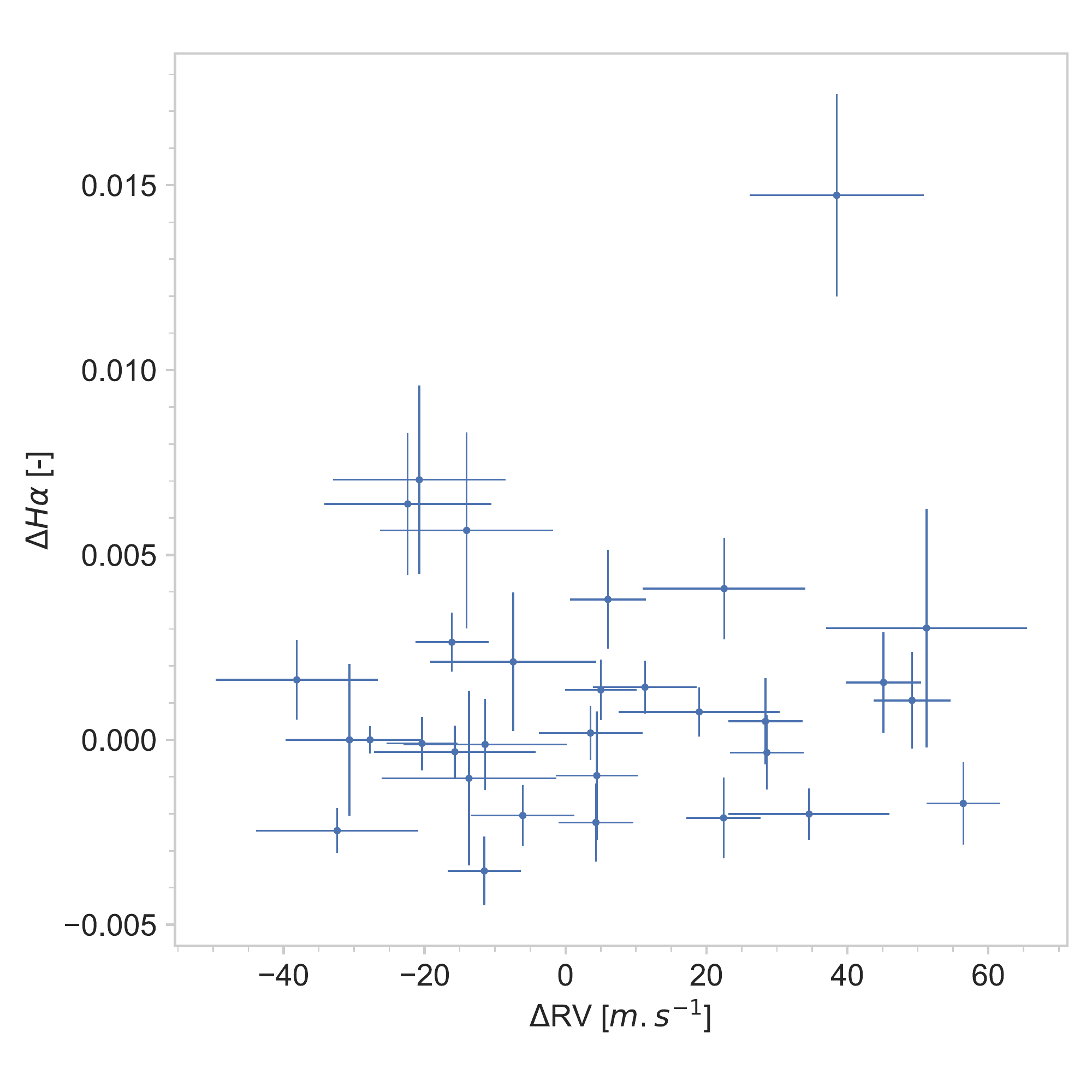}
        \caption{Correlation plot of the radial velocities and H$\alpha$ activity index time-series for HD\,29399. A non-significant correlation is observed, with a weighted Pearson coefficient value of $R_{P}=0.078\pm0.103$ and a weighted Spearman's rank of $R_S=0.017\pm0.106$.}
        \label{fig:corr_rv_ha_index_hd29399}
\end{figure}

Our standard approach to search for periodic signals in radial-velocity time-series is to follow an iterative process consisting in looking for successive significant dominant peaks in the periodogram of the corresponding radial-velocity residuals. At each step of the iteration, the radial-velocity residuals are computed by readjusting the full model composed of the N independent Keplerians, potential linear, quadratic, or cubic drift terms to fit long-term trends, the different instrumental offsets, and additional white noise terms\footnote{We fit a combination of white-noise terms corresponding to individual instrumental precision and intrinsic stellar jitter. The instrumental precision is well known for each version of CORALIE: $\sigma_{COR98}=5.0\pm0.5$\,m\,s$^{-1}$, $\sigma_{COR07}=8.0\pm0.5$\,m\,s$^{-1}$, $\sigma_{COR14}=3.0\pm0.5$\,m\,s$^{-1}$}. We proceeded with the periodicity search by computing the periodogram of the data in the range 10$-$10\,000\,days using the algorithm implemented on DACE \citep[see][]{Delisle2020a,Delisle2020b}, and using the false alarm probability (FAP) to assess the significance of the signal following the formalism of \citet{Baluev2008}. 

A periodic signal at 896 days clearly stands out in the periodogram of the radial velocities of HD\,29399 (\autoref{fig:spec_perio}), but the signal could  have various origins. Radial-velocity variations can be linked to several, potentially periodic effects that can mimic planetary signals. Giant stars exhibit short-period solar-like radial pulsations \citep{Walker1989,Hatzes1993,Hatzes1994,Frandsen2002,DeRidder2006,Hekker2006a}, as well as non-radial oscillations \citep{Hekker2006c,DeRidder2009,Hekker2010b} with  lifetimes of hundreds of days \citep{Dupret2009}. They can also produce longer period variations from a combination of magnetic cycles \citep{Santos2010,Dumusque2011a}, beating of modes, or rotational modulations of features on the stellar surface (starspots, granulation, etc.). To test those possible sources of radial-velocity periodic signal, we also thoroughly checked for variations of the line profile of the cross correlation function (CCF), and of the classical spectroscopic chromospheric indicator (in this case the H$\alpha$ index), as well as the long-term photometric variation of the star. This is described in the following sections.

\subsubsection{Spectroscopic indicator variations}\label{subsub:spectro_indic}

Intrinsic stellar variability can be tracked through changes in the shape of the spectral lines, and consequently of the CCF (the product of convolution of the spectrum with a template, used for the estimate of the radial velocities). The profile of the CCF can be monitored by computing its first moments \citep{Aerts2000}. Here, as proxies, we use the contrast, the full width at half-maximum (FWHM), and the bisector inverse span (BIS) measured on the CCF. In the case of HD\,29399, none of these show any significant periodicity in their respective periodogram (\autoref{fig:spec_perio}). We also checked for potential correlations between these parameters and radial velocities by computing the corresponding weighted Pearson coefficient. No significant correlation was found.

On the other hand, we have to mention that \citet{Wittenmyer2017a} reported a $765$-day variation signal with a 50\,m\,s$^{-1}$ semi-amplitude in the radial-velocity data obtained with the UCLES and CHIRON spectrographs, which they associated with an intrinsic stellar variation. After checking for several potential `intrinsic' origins for the variation,  inconclusively\footnote{Spots were rejected because of the large surface coverage required for HD\,29399. Such coverage is needed because of the slow rotation and large radius that lead to a lower estimate of the period of rotation of $\sim$169\,days. The presence of a debris disc, as first hypothesised by \citet{Wittenmyer2017a}, was also excluded because to explain the radial-velocity variation it would have required the debris disc to be heated to 1500\,K, which is too close to the star to produce such a long periodicity}, the authors observed correlations between the radial velocities and photometric data, and the equivalent width of H$\alpha$ activity (see their Fig.\,6), which led them to the conclusion that the radial-velocity signal was intrinsic to the star, and the `planet' labelled a false positive.


To check this affirmation, we monitored the impact of magnetic activity on the chromosphere by computing the H$\alpha$ chromospheric index from our CORALIE spectra, produced the same figure as in \citet{Wittenmyer2017a}, and computed the weighted Pearson correlation coefficient R$_P$ and the Spearman's rank R$_S$ using a bootstrap randomisation technique. These estimates lead to non-significant correlation values: R$_P=0.078\pm0.103$ and R$_S=0.017\pm0.106$ (see \autoref{fig:corr_rv_ha_index_hd29399}). We therefore do not confirm the result reported in \citet{Wittenmyer2017a}.


\subsubsection{Photometric variability analysis}\label{subsub:photo_var}
\citet{Wittenmyer2017a} also checked  the All-Sky Automated Survey V-band photometric data \citep[ASAS-3,][] {Pojmanski2002} for variability due to intrinsic stellar processes, and a $\sim$765-day periodicity was also found in the data \citep[see][Fig.\,1]{Wittenmyer2017a}. This highly significant signal and the large photometric variability of the star further supported the interpretation of the radial-velocity signal as being intrinsic to the star.

Following \citet{Wittenmyer2017a} we also checked the ASAS-3 data \footnote{Courtesy of Grzegorz Pojmanski, who sent us the complete photometric dataset, along with technical information on the survey and precautions to take when using the data.} for variability due to intrinsic stellar processes or surface rotational modulation. We only considered the data flagged as \textit{GRADE A} quality. It is important to point out that the ASAS-3 photometry has a saturation level at $\sim$6\,mag, depending on the observing conditions and epoch, which clearly impacts the consistency and quality of the data acquisition for our star. The focus of the camera was not stable over time because of many instrumental issues, the observing scheme and data handling pipeline have evolved over the years although the data have never been reduced again, and there are obvious consequences of varying weather conditions such as extinction due to tiny clouds, increased seeing due to wind, fog, and tracking problems.


From 1850 to 2400 [HJD-2\,450\,000], there was no treatment of saturation, causing important discontinuities observed in the data, and exposures were 180\,s in length. The induced patterns can generate non-negligible signals in the periodogram of the data. Later, some saturation corrections were included. After 2950 [HJD-2\,450\,000] (Nov, 2003), instead of taking one 180\,s exposure, the standard observation consisted of three 60\,s exposures, which reduced the saturation level by over 1\,mag. From that point until the end of 2010, the system was considered as more or less stable. Despite the fact that the saturation correction procedure improved the data, for bright stars (V\,<\,$\sim~$6.5-7) saturation is still present and is a clear concern for HD\,29399. We also observe in several targets of our sample that the photometric data between 2950 and 3300 [HJD-2\,450\,000] presents a very low dispersion clump that does not match the data acquired later. This `step' creates the equivalent of a long period trend that completely disappears when removing this portion of the data.

\begin{figure}[t!]
        \centering
        \adjincludegraphics[width=1\columnwidth, trim={0 0 0 0},clip]{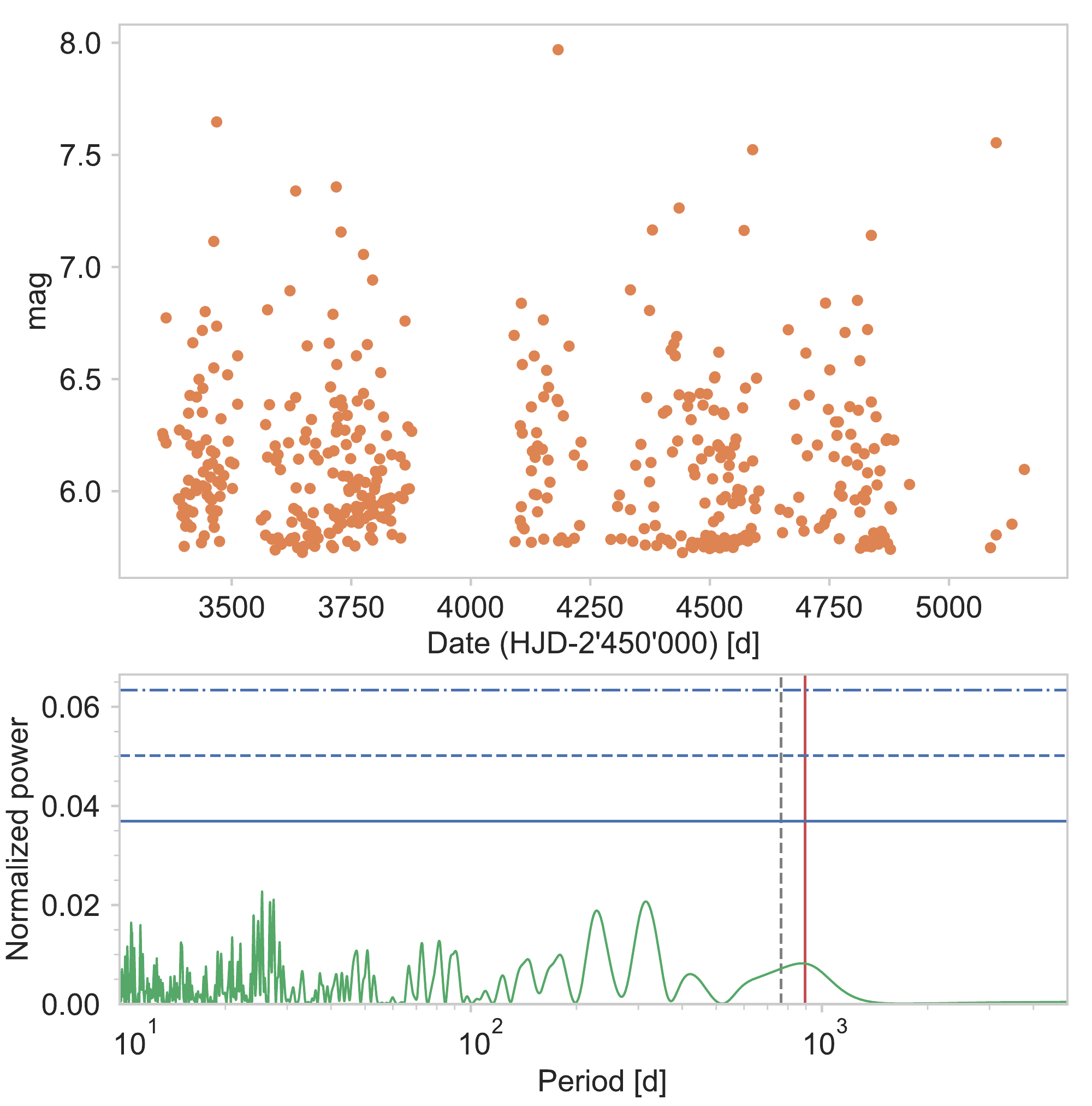}
        \caption{Light curve (top) and periodogram (bottom) of HD\,29399 from the ASAS-3  data (\citet{Pojmanski2002}). (Top) Data acquired after 3300 [HJD-2\,450\,000], the ASAS-3300 dataset (see text). (Bottom) Periodogram of the ASAS-3300 dataset. The red vertical line corresponds to the 897.2-day period found in the radial velocities, and the dashed-vertical line to the 765-day period found by \citet{Wittenmyer2017a}. Horizontal lines, from bottom to top, are the FAP levels at 10\%, 1\%, and 0.1\% respectively.}
        \label{fig:asas_hd29399}
\end{figure}

\begin{table}[t]
\centering
\begin{threeparttable}
\caption{Radial-velocity observation statistics, best-fit solutions of the model with instrumental offsets, nuisance parameters, Keplerian orbital parameters, and inferred planetary parameters.}
    \begin{tabular}{llc}        
    \hline
    \hline
    & & HD\,29399b \\
    \hline
    \multicolumn{3}{c}{Observations}\\
    \hline
    $N_{obs}$ & & 61 \\
    $T_{span}$ & $[days]$ & 4811 \\
    $rms_{tot}$ & $[m.s^{-1}]$ & 24.71 \\
    $rms_{res}$ & $[m.s^{-1}]$ & 11.18 \\
    $\chi^2_{red}$ & & 1.46 \\
    \hline
    \multicolumn{3}{c}{Offsets $^{(1)}$}\\
    \hline
    $\gamma_{COR07}$ & $[m/s]$ & 31659.7~$\pm$~3.0 \\
    $\Delta\,RV_{COR14-COR07}$ & $[m/s]$ & 12.2~$\pm$~3.7 \\
    $\Delta\,RV_{CHIRON-COR07}$ & $[m/s]$ & -31641.5~$\pm$~6.9 \\
    $\Delta\,RV_{UCLES-COR07}$ & $[m/s]$ & -31663.0~$\pm$~3.8 \\
    $\Delta\,RV_{HARPS03-COR07}$ & $[m/s]$ & 20.6~$\pm$~6.8 \\
    \hline
    \multicolumn{3}{c}{Instrumental Noises}\\
    \hline
    $\sigma_{COR98}$ & $[m/s]$ & 5.3~$\pm$~1.0 \\
    $\sigma_{COR07}$ & $[m/s]$ & 7.8~$\pm$~1.4 \\
    $\sigma_{COR14}$ & $[m/s]$ & 3.0~$\pm$~0.5 \\
    $\sigma_{CHIRON}$ & $[m/s]$ & 7.6~$\pm$~7.1 \\
    $\sigma_{UCLES}$ & $[m/s]$ & 4.7~$\pm$~3.3 \\
    $\sigma_{HARPS03}$ & $[m/s]$ & 3.4~$\pm$~2.6 \\
    \hline
    \multicolumn{3}{c}{Stellar Jitter}\\
    \hline
    $\sigma_{jit}$ & $[m.s^{-1}]$ & 8.9~$\pm$~1.5 \\
    \hline
    \multicolumn{3}{c}{Drifts}\\
    \hline
    Lin. & $[m/s/yr]$ & 1.9~$\pm$~0.6 \\
    \hline
    \multicolumn{3}{c}{Keplerians}\\
    \hline
    $P$ & $[days]$ & 892.7~$\pm$~5.9 \\
    $K$ & $[m.s^{-1}]$ & 29.9~$\pm$~2.2 \\
    $e$ & & 0.05~$\pm$~0.05 \\
    $\omega$ & $[deg]$ & -13.1~$\pm$~85.1 \\
    $\lambda_0$ $^{(2)}$ & $[deg]$ & 151.9~$\pm$~5.0 \\
    $T_p$ $^{(2)}$ & $[rjd]$ & 5965.0~$\pm$~210.0 \\
    \hline
    $a$ & $[au]$ & 1.913~$\pm$~0.008 \\
    $m_2\,sin\,i$ $^{(3)}$ & $[M_J]$ & 1.57~$\pm$~0.11 \\    
    \hline
    \end{tabular}
\begin{tablenotes}
    \small 
    \item {$^1$} The reference instrument is COR07.
    \item {$^2$} The mean longitude is given at $BJD=2\,450\,000$ [d] while $2\,450\,000$ has been subtracted from the date of passage through periastron (T$_P$).
    \item {$^3$} Using the model-independent mass from seismic inversions (see Sect. \ref{SecModelling})
\end{tablenotes}
\label{tab:orbit_params}
\end{threeparttable}
\end{table}

\begin{figure}[btp]
        \centering
        \adjincludegraphics[width=1\columnwidth, trim={{.03\width} {.3767\height} 0 0},clip]{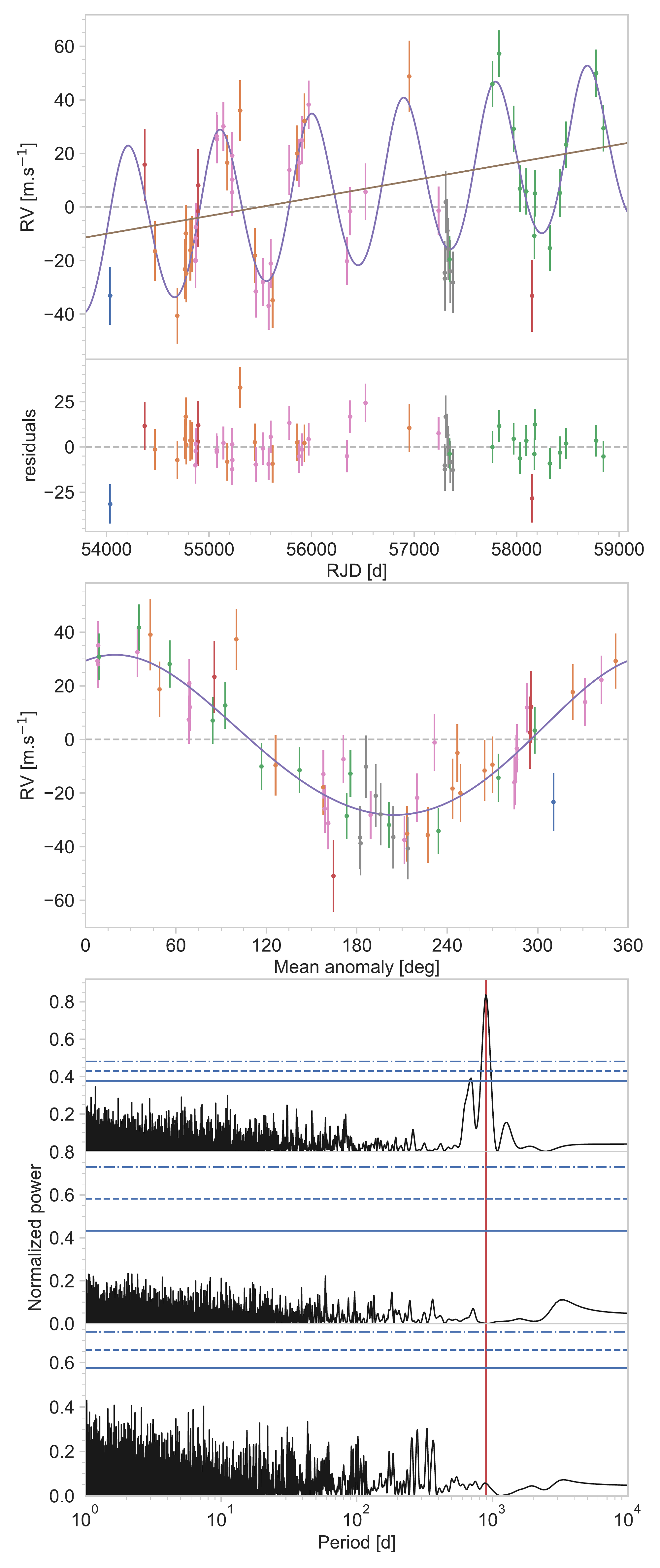}
        \caption{First panel: Radial velocities of HD\,29399 from CORALIE (COR98 in blue, COR07 in orange and COR14 in green), HARPS (in red), UCLES (in pink), and CHIRON (in grey). Overplotted are the fitted Keplerian orbit (purple curve) and the linear trend (dark line). Second panel: Residual radial velocities after subtracting the Keplerian fit. Third panel: Radial velocities phased to the fitted period at 890.9 days. We note that the uncertainties of the HARPS point have been increased to take into account the calibration of the offset and the intrinsic variability of the star (see text).}
        \label{fig:timeseries_hd29399}
\end{figure}

For these reasons and to be on the safe side, we decided to only use for our analysis the data acquired after 3300 [HJD-2\,450\,000] (which we refer to as \textit{`ASAS-3300'}) and periodograms, as shown in \autoref{fig:asas_hd29399}. The \textit{ASAS-3300} dataset contains 466 epochs spanning 4.9 years with a mean value of 6.12\,$\pm$\,0.05\,mag and 0.35\,mag rms. The dataset has a span of twice the periodicity found in our CORALIE radial-velocity time-series, and is thus still completely adequate to check for stellar photometric variability. For the analysis, we also systematically removed outliers from this subset following a sigma-clipping process. We observe that those outliers were usually fainter, which reinforces our intuition that they are linked to bad weather conditions.

The periodogram of the complete photometric time-series shown in Fig.1 of \citet{Wittenmyer2017a} clearly exhibits signals around 200\,days (contribution of data acquired between 1850 and 2400 [HJD-2\,450\,000]) and 765\,days (contribution of data acquired between 2400 and 3300 [HJD-2\,450\,000]) that completely disappear in the periodogram of the more reliable \textit{ASAS-3300} data (see \autoref{fig:asas_hd29399}). We can therefore conclude that the photometric analysis from \citet{Wittenmyer2017a} suffered from the non-optimal quality of the early ASAS data. The conclusion of a false-positive planetary signal in the radial velocities caused by features on the surface of the star is therefore not robust.


\section{Keplerian analysis of the radial velocities}\label{sec:kep_analysis}
In the absence of any significant periodic signal and correlations in the activity-related products from the high-resolution spectra, we assumed that the periodic variation observed in the radial-velocity time-series of HD\,29399 is due to a substellar companion orbiting the star. Following the procedure described in \autoref{subsec:hd29399_obs}, we detected a dominant peak in the periodogram at a period of $\sim$890.91\,days corresponding to a variation with a semi-amplitude of $\sim$30\,ms$^{-1}$. To fit the parameters of the model, we used the MCMC algorithm implemented in DACE, developed by \citet{Diaz2014,Diaz2016}, to probe the complete parameter space, with $1.6$ million iterations. We used the following parameters for the Keplerian model: We used the natural logarithm of the period (log\,P) and of the semi-amplitude (log\,K) to better explore ranges of several orders of magnitude with a uniform prior; $\sqrt{e \cos{\omega}}$ and $\sqrt{e \sin{\omega}}$ (with $e$ the eccentricity of the orbit and $\omega$ the argument of the periastron) to obtain a uniform prior for the eccentricity; and finally the mean longitude at epoch of reference (i.e., $BJD=2\,455\,500$ [d]) ($\lambda_0$), with a uniform prior. We used a uniform prior for the COR07 offset of reference, and Gaussian priors for the relative offsets between COR07 and COR98/14: $\Delta\,RV_{COR98-COR07}$: $\mathcal{N}(0,4)$ m\,s$^{-1}$, $\Delta\,RV_{COR14-COR07}$: $\mathcal{N}(14,4)$ m\,s$^{-1}$. We also used Gaussian priors for the instrumental noise: $\sigma_{COR98}$: $\mathcal{N}(5,1)$ m\,s$^{-1}$, $\sigma_{COR07}$: $\mathcal{N}(8,1.5)$ m\,s$^{-1}$ and $\sigma_{COR14}$: $\mathcal{N}(3,0.5)$ m\,s$^{-1}$. Finally, we used a uniform prior for the stellar jitter parameter. We present in \autoref{apdx:corner} the corner plot of the posterior distributions of the fitted parameters. Adopting the model-independent stellar mass of 1.17\,M$_{\odot}$ (see \autoref{SecModelling}), the single-planet model yields a minimum mass for the companion of 1.57\,M$_{J}$, on a 892.7-day period orbit with a semi-major axis of 1.91\,au.

In addition, HD\,29399 exhibits a clear long-term radial-velocity trend, probably explained by the presence of an additional companion. The observed slope could correspond to a substellar object of a couple of Jupiter masses, with a minimum period of $\sim$8000\,days, considering a circular orbit. A longer time-span is necessary to better constrain this trend. It can be adequately fitted by a linear drift term, however part of the effect could be related to the effect of the offset between the disjointed CORALIE instruments. The use of the four HARPS points\footnote{The last HARPS point was obtained after a major intervention on the instrument, changing the instrument zero point. An offset was then applied (based on \citet{LoCurto2015}). We also increased the uncertainties of the HARPS points to take into account the effect of the calibration and the minimum intrinsic variability of the star.} broadly spread in time, as well as the inclusion of the CHIRON and UCLES data sets overlapping with the CORALIE measurements, help to consolidate the estimate of the offsets between the instruments. In this context, we tried three types of models: a single planet, a planet $+$ a linear drift, and a planet $+$ a quadratic drift. A Bayesian model comparison using the Bayesian information criterion clearly indicates a preference for the model with one planet $+$ linear drift (change in the BIC of 10 and 2, adding a linear drift, and then a quadratic term, respectively).    

The resulting best model overplotted on the radial velocities, as well as the residuals around the best solution are shown in \autoref{fig:timeseries_hd29399}. Table~\ref{tab:orbit_params} presents the statistics of the distributions (i.e., the median and standard deviation) of the most common set of Keplerian parameters P, K, e, $\omega$, and the date of passage through periastron (T$_P$), as well as the distributions of the semi-major axis and minimum masses derived from the MCMC chains of the fitted parameters. \Autoref{apdx:corner} shows the corner plot of the posterior distributions of the fitted parameters. The weighted rms of the residuals around the solution is comparable to the radial-velocity dispersion of giant stars with similar $B-V$ (see Fig.3 from \citealt{Hekker2006b}).

In light of the presented results, we are fairly confident that the 892.7-day periodic variation in the radial-velocity data of HD\,29399 is not due to chromospheric stellar activity as there are no significant correlations between the radial velocities and the classical chromospheric indicators. Rotational modulation of surface features such as spots can also be excluded, as they would require a very large percentage of the stellar surface to be covered, and we do not observe any long-term photometric variability. Finally, we found no trace of long-period non-radial oscillation modes (either matching periodicities or corresponding to harmonics in the line profile moments). An orbiting planet therefore remains a valid hypothesis to explain the radial-velocity periodic signal detected.

Finally, it is worth noting that long-period planets orbiting giant stars are good candidates for transit search. Indeed, the decrease in transit probability due to the long period is  compensated by the large radius of the star. The only downside to using such targets is the small depth of the transit scaling with the square of the star-to-planet ratio. However, this remains detectable from space with missions such as TESS \citep{Ricker2015}, CHEOPS \citep{Benz2020}, and PLATO \citep{Rauer2014}. For the planet orbiting HD\,29399, the transit probability is of the order of 1\,\% and an expected depth of $\sim500$\,ppm for a Jupiter-size planet.

\section{Asteroseismic analysis}\label{SecAstero}

\subsection{Peakbagging and frequency determinations}\label{AsteroObs}
We studied the acoustic oscillation mode frequencies of HD 29399 (also known as TIC 38828538) using photometric measurements from TESS. Our method comprises constructing a power spectrum from the observed flux and measuring the locations of radial $(\ell=0)$ and quadrupolar $(\ell=2)$ oscillation modes using the automated peakbagging package \texttt{PBjam}\footnote{See \url{https://github.com/grd349/PBjam}} \citep{Nielsen2020}.

To construct the power spectrum, we used the \texttt{lightkurve} package \citep{lightkurvecollaborationLightkurveKeplerTESS-2018} which makes use of the \texttt{astropy} \citep{AstropyCollaboration.Robitaille.ea2013,AstropyCollaboration.Price-Whelan.ea2018} and \texttt{astroquery} packages \citep{Ginsburg.Sipocz.ea2019}. We downloaded TESS light curves from the Mikulski Archive for Space Telescopes (MAST) for all sectors from 1 to 12, except sector 3. Subsequently, we stitched the Pre-search Data Conditioning Simple Aperture Photometry \citep[PDCSAP,][]{Stumpe.Smith.ea2012,Smith.Stumpe.ea2012} flux for each sector, removing both low-frequency trends, using a Savitzky–Golay filter, and $5$-$\sigma$ outliers. We obtained the power spectrum using the Lomb-Scargle method \citep{Lomb1976,Scargle1982}. To determine the signal-to-noise ratio, we divided the power spectrum by an estimate of the background obtained by smoothing with a moving median in steps of $\frac{1}{2}\log(0.01\,\mu\mathrm{Hz})$.

We used the \texttt{PBjam} package to determine the observed radial, $\nu_{n,0}$, and quadrupolar, $\nu_{n,2}$, oscillation modes of the star. Initial mode identification was based on the methods of \citet{daviesAsteroseismologyRedGiants-2016} and a prior probability distribution constructed from thousands of stars already analysed using \texttt{PBjam}. The means, $\mu$, and uncertainties, $\sigma$, on the input parameters used to select stars from the prior are given in \autoref{tab:seismo_input}. The input large frequency separation, $\Delta\nu$, and frequency at maximum power, $\nu_{\mathrm{max}}$, were obtained using the methods described in \citet{hekkerSolarlikeOscillationsRed-2012}. We adopted the input effective temperature, $T_{\mathrm{eff}}$, and colour, $G_{\mathrm{BP}} - G_{\mathrm{RP}}$, from \textit{Gaia} DR2 \citep{GaiaCollaboration.Prusti.ea2016,GaiacollaborationGaiaDataRelease-2018}. The inputs primarily determined the window in which we selected stars from the prior for subsequent mode identification; they had no influence on the final peakbagging step.

We performed initial mode identification by fitting the asymptotic relation \citep{2013A&A...550A.126M} and used the \texttt{emcee} package \citep{foreman-mackeyEmceeMCMCHammer-2013} to build the final posterior distributions of the modes and provide their identification. This resulted in values for the large frequency separation, $\Delta\nu = 14.93 \pm 0.02\,\mu\mathrm{Hz}$, and the frequency at maximum power, $\nu_{\mathrm{max}} = 196.5 \pm 2.5\,\mu\mathrm{Hz}$, both within $1$- and $2$-$\sigma$ of their input values, respectively.

\begin{table}
    \begin{center}
    \caption{Global stellar property estimates for TIC 38828538 used as inputs for the \texttt{PBjam} peakbagging pipeline.}
    \label{tab:seismo_input}
    \begin{tabular}{ccc}
    \hline \hline
    Input & $\mu$ & $\sigma$ \\
    \hline
    $\Delta\nu\,(\mu\mathrm{Hz})$ & 14.88 & 0.05 \\
    $\nu_{\mathrm{max}}\,(\mu\mathrm{Hz})$ & 193.34 & 1.47 \\
    $T_{\mathrm{eff}}\,(\mathrm{K})$ & 4803 & 70 \\
    $G_{\mathrm{BP}} - G_{\mathrm{RP}}\,(\mathrm{dex})$ & 1.17 & 0.01 \\
    \hline
    \end{tabular}
    \end{center}
\end{table}
        
\begin{table}
    \begin{center}
    \caption{Individual observed asteroseismic frequencies for TIC 38828538. The radial and quadrupolar oscillation modes, with their statistical uncertainties, are given in the left and right columns respectively.}
    \label{tab:seismo_output}
    \begin{tabular}{cc}
    \hline \hline
    $\nu_{n,0}\,(\mu\mathrm{Hz})$ & $\nu_{n,2}\,(\mu\mathrm{Hz})$ \\
    \hline
    $139.619\pm0.070$ & $137.717\pm0.257$ \\
    $154.063\pm0.051$ & $152.152\pm0.061$ \\
    $168.590\pm0.027$ & $166.727\pm0.043$ \\
    $183.575\pm0.024$ & $181.751\pm0.043$ \\
    $198.352\pm0.017$ & $196.522\pm0.035$ \\
    $213.258\pm0.040$ & $211.439\pm0.031$ \\
    $228.564\pm0.039$ & $226.817\pm0.032$ \\
    $244.017\pm0.224$ & $241.788\pm0.240$ \\
    \hline
    \end{tabular}
    \end{center}
\end{table}
    
After the mode identification, we fit a Lorentzian profile to each mode by sampling its posterior distribution using the Bayesian package \texttt{PyMC3} \citep{Salvatier.Wiecki.ea2016}. Each mode location, $\nu_{n,l}^{\,\prime}$, from the previous step was used as the mean of a prior normal distribution, $\mathcal{N}(\mu, \sigma)$, on the Lorentzian centre given by,
\begin{equation}
    \nu_{n,l} \sim \mathcal{N}\left(\nu_{n,l}^{\,\prime}, 0.03\Delta\nu\right),
\end{equation}
where $\mu$ and $\sigma$ are the mean and standard deviation, respectively. All other parameters from the previous steps were relaxed.

\begin{figure}
\centering
    \adjincludegraphics[width=1\columnwidth, trim={0 0 0 0},clip]{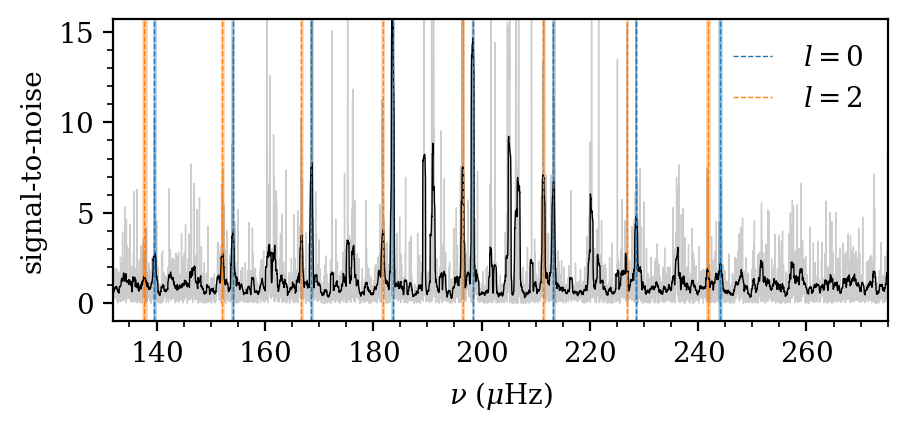}
    \caption{Asteroseimic signal-to-noise ratio power spectrum for HD\,29399 (TIC 38828538) in light grey and the smoothed spectrum in black. The locations of the radial, $l=0,$ and quadrupolar, $l=2,$ oscillation modes are shown with dashed lines, and their 68\% credible regions are shaded in blue and orange, respectively.}
    \label{fig:seismo_model}
\end{figure}

\begin{figure}
\centering
    \adjincludegraphics[width=1\columnwidth, trim={0 0 0 0},clip]{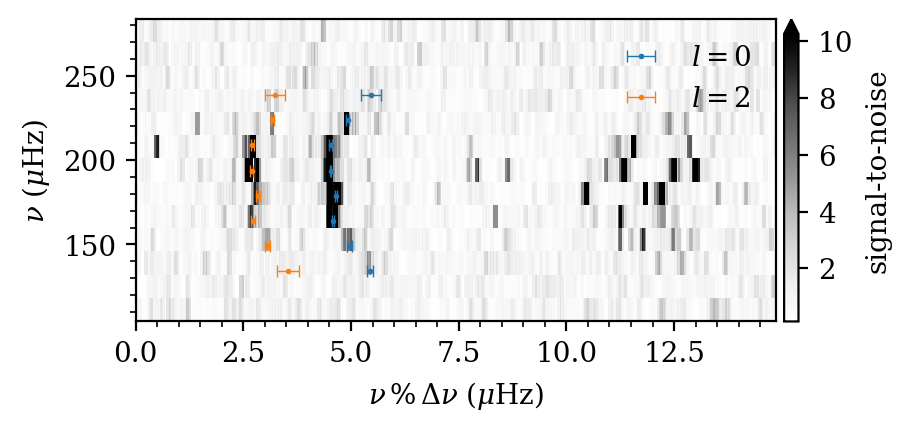}
    \caption{Echelle diagram for HD\,29399 (TIC 38828538). The signal-to-noise ratio in frequency, $\nu,$ modulo the large frequency separation, $\Delta\nu = 14.93\,\mu\mathrm{Hz}$. The locations of radial, $l=0$ (blue), and quadrupolar, $l=2$ (orange), oscillation modes are shown with error bars given by their 68\% credible regions.}
    \label{fig:seismo_echelle}
\end{figure}

The oscillation mode locations are given in \autoref{tab:seismo_output}, and plotted on the periodogram in \autoref{fig:seismo_model} together with their $68\%$ credible regions. These are the observed frequencies, uncorrected for any shifts due to the radial-velocity of the star. The echelle diagram in \autoref{fig:seismo_echelle} shows the locations of the radial and quadrupolar oscillation modes phase-folded by the large frequency separation determined above. We give the uncertainties as the standard deviation of the posterior samples for each mode location.

\subsection{Seismic modelling}\label{SecModelling}

In this section, we carry out a detailed modelling of HD\,29399, combining seismic and classical constraints from both ground-based spectroscopic surveys and \textit{Gaia} parallax values. We use the Liège stellar evolution code \citep{ScuflaireCles} combined with the Liège stellar oscillation code \citep{ScuflaireOsc} to compute adiabatic oscillations. The available constraints are summarised in \autoref{tab:stellar_params} with their corresponding references. Following \citet{Buldgen2019Kepler}, the modelling is divided into three steps:
\begin{enumerate}
\item Forward modelling with the Asteroseismic Inference on a Massive Scale \citep[AIMS,][]{Rendle2019,Montalban2020} software using radial oscillations and classical constraints.
\item Inversion of the stellar mean density to determine a model-independent mass range.
\item Forward modelling combining the inversion results, classical constraints, and frequency separations of the radial and quadrupolar modes. 
\end{enumerate}

The first step uses a global minimisation technique to determine first estimates of the global stellar properties and carry out a thorough exploration of the parameter space. This approach ensures a reliable and accurate inversion procedure in step 2, which is used to determine the model-independent mass interval from which the planetary properties are deduced. Using these results, we carry out a third step using a local minimisation technique to find a solution in better agreement with the observations, delivering a robust age estimate for the system better accounting for seismic and non-seismic constraints.

In the first step, we used two different grids of stellar evolutionary models described in \citet{Rendle2019} and used in \citet{Buldgen2019} to study a sample of \textit{Kepler} eclipsing binaries. This grid uses the GN93 solar abundances \citep{GrevNoels} and the corresponding metallicity scale. We also recomputed a second grid using the AGSS09 revision of the solar abundances \citep{AGSS09} to explore their impact on the determined stellar parameters. Both grids are evolved up to a cut-off lower $\nu_{\rm{Max}}$ value determined in \autoref{AsteroObs}. Moreover, \citet{Rendle2019} and \citet{Buldgen2019} used an Eddington atmosphere for the outer boundary layers, which is unsuitable for fitting classical parameters on the RGB with a solar-calibrated mixing-length parameter value. Therefore,  in the second grid we used a $T(\tau)$ relation from Model C of \citet{Vernazza}, which is more suitable for our needs \citep{Sonoi2019}. None of the grids include microscopic diffusion. The observed frequencies have been corrected from the line-of-sight Doppler velocity shifts following the recommendations of \citet{Davies2014}. Both grids assume a solar calibrated mixing-length parameter, use the FreeEOS equation of state \citep{Irwin}, and OPAL opacity tables \citep{OPAL}. We summarise the properties of the AGSS09 grid in \autoref{tabGridProp} and refer the reader to \citet{Rendle2019} for the properties of the GN93 grid. 

\begin{table}
\caption{Properties of the AIMS stellar evolution model grids.}
\label{tabGridProp}
  \centering
\begin{tabular}{|r | c | c}
\hline \hline
Parameters& AGSS09 grid values \\ \hline
Mass $\left( M_{\odot} \right)$&$1.00-1.70 $ $(0.02\; \rm{step})$\\
$X_{0}$ &$\left[0.68, 0.72 \right]$ $(0.01\; \rm{step})$\\
$Z_{0}$&$\left[ 0.010, 0.050 \right]$ $(0.001\; \rm{step})$\\
$\alpha_{\rm{MLT}}$ $\left(H_{P}\right)$&$2.03$\\
$\nu_{\rm{Max}}$ cutoff $\left( \mu \rm{Hz}\right)$&$40$\\
\hline
\end{tabular}

\end{table}

\begin{figure}[t]
\centering      
        \adjincludegraphics[width=.49\columnwidth, trim={95 32 60 93}, clip]{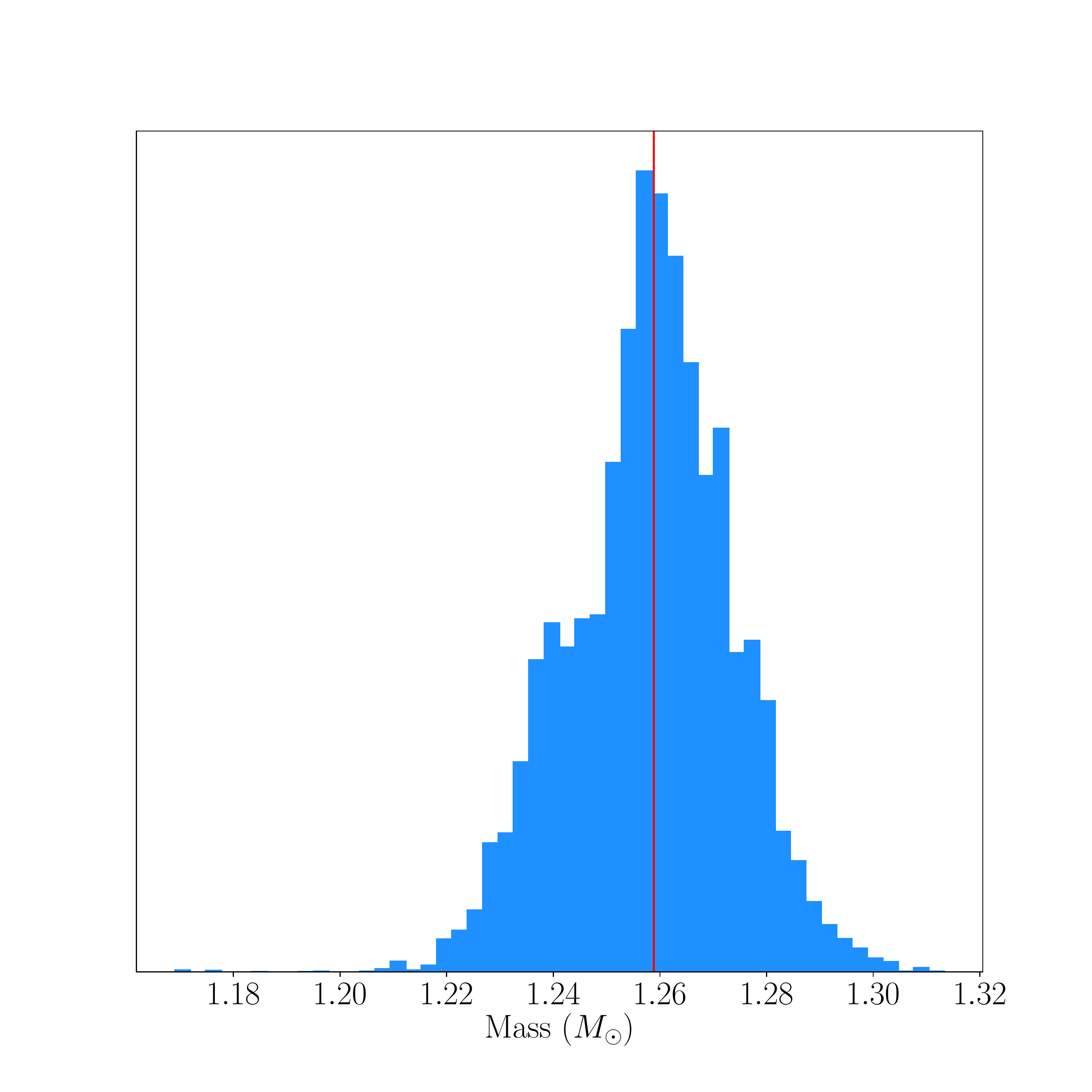}
        \adjincludegraphics[width=.49\columnwidth, trim={95 32 60 93}, clip]{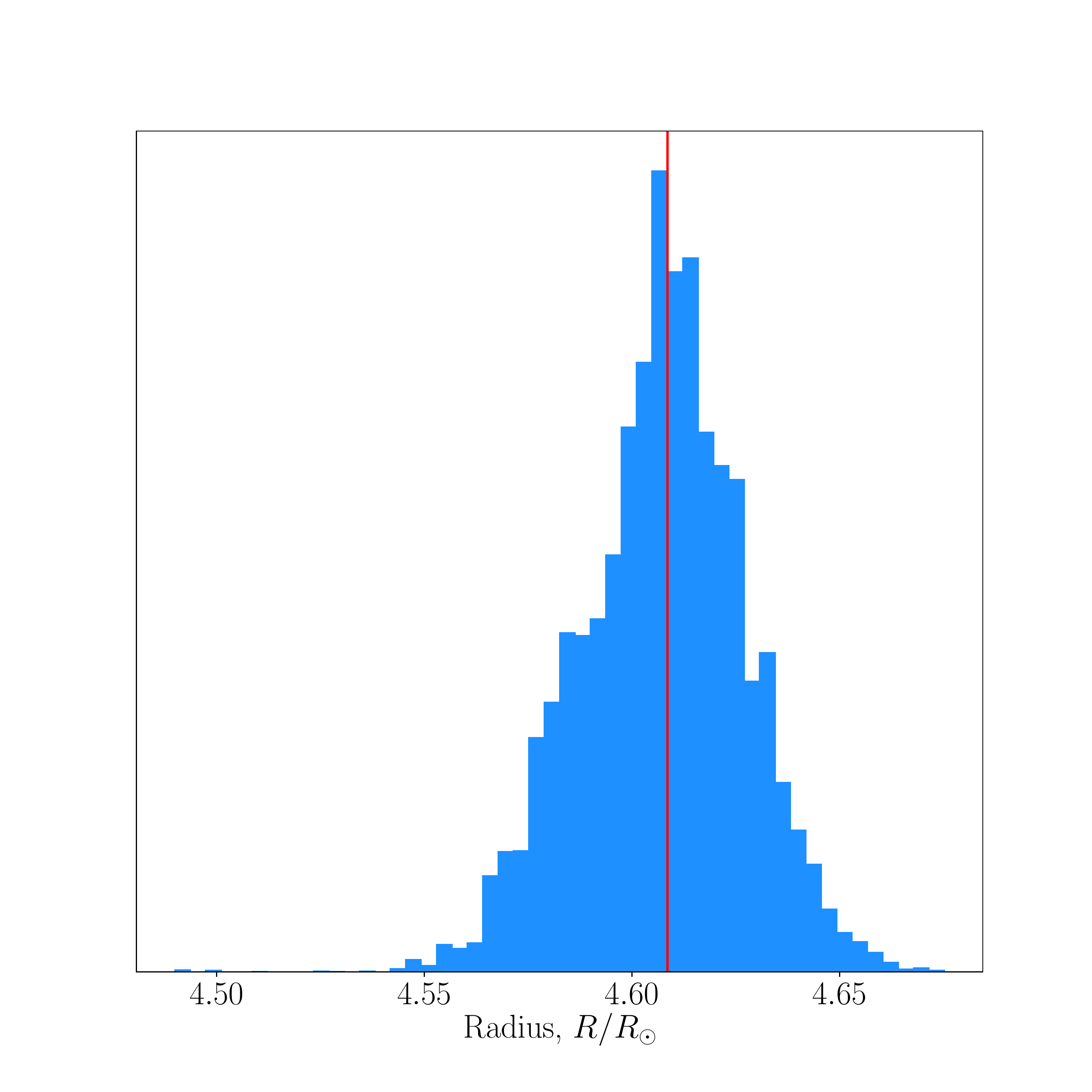}
        \caption{Probability distribution functions for the mass (left) and radius (right) for HD\,29399 (TIC38828538) obtained using AIMS. The vertical lines indicate the position of the best model obtained from a simple scan of the grid.}
        \label{FigDistrib}
\end{figure}

\begin{figure}
        \centering
        \adjincludegraphics[width=.8\columnwidth, trim={0 0 0 {0.021\height}},clip]{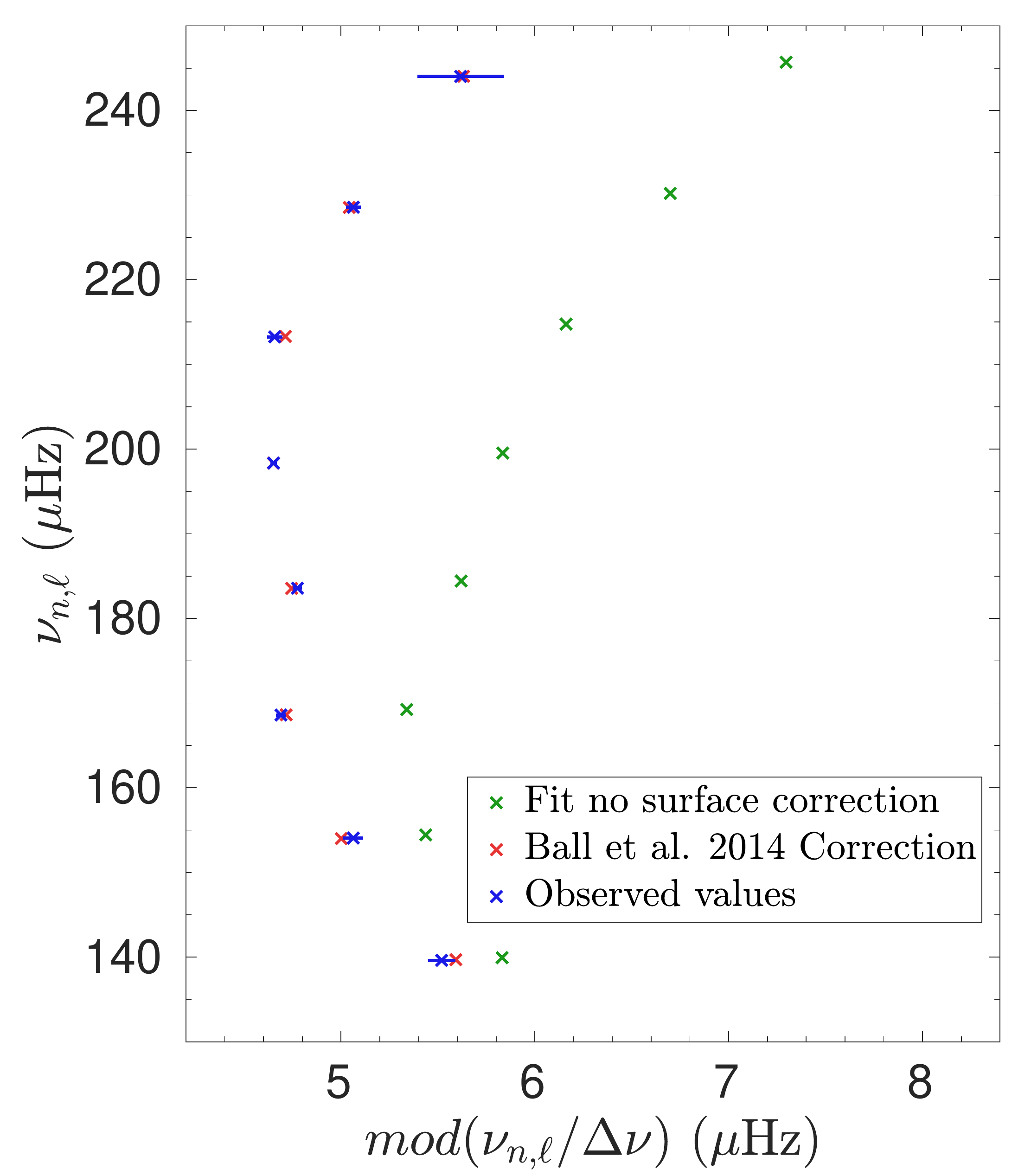}
        \caption{Echelle diagram illustrating the agreement between theoretical and observed radial $(\ell=0)$ frequencies for the AIMS AGSS09 solution.}
                \label{FigEchAIMS}
\end{figure}

The results are illustrated for the AGSS09 grid in \autoref{FigDistrib}. As we can see, the individual radial frequencies are relatively well fitted by AIMS. A similar conclusion is reached for the final $\left[ \rm{Fe}/\rm{H} \right]$ of the model, which is $0.095 ~\rm{dex,}$ and the effective temperature of $4880~K$. However, the radius of the model slightly disagrees with the value determined from \textit{Gaia} and the spectroscopic constraints reported in \autoref{tab:stellar_params}. This small discrepancy is also seen in the luminosity of the model, found around $10.78~L_{\odot}$ and results from the higher weight of the individual frequencies compared to the classical constraints. Nevertheless, these discrepancies remain very small and within $1\sigma$ of the observations. The age of the star is found to be around $\rm 6.33 ~Gyr$. The results for the GN93 grid are similar, with good agreement found in frequencies, but, unsurprisingly given the atmosphere used, significant mismatches are found for the classical parameters. Nevertheless, given that they fit the individual frequencies (as shown in \autoref{FigEchAIMS}) for the AGSS09 model, both models are suitable for seismic inversion. 

Following \citet{ReeseDens} and \citet{Buldgen2019}, we carried out an inversion of the mean density using only radial modes to avoid non-linear behaviours. The inversion procedure is based on the integral relations linking relative frequency differences to corrections of thermodynamic variables of the stellar structure \citep{Dziemboswki90}. From these relations, we can determine the coefficients, $c_{i}$, such that the mean density of the reference model is computed in a model-independent way from a recombination of the individual frequencies:
\begin{align}
\frac{\delta \bar{\rho}}{\bar{\rho}}=\sum_{i}c_{i}\frac{\delta \nu_{i}}{\nu_{i}}.
\end{align}

We used the SOLA inversion technique \citep{Pijpers}, following the guidelines of \citet{ReeseDens} and \citet{Buldgen2019}, applying different surface effect corrections \citep[namely those of][]{Kjeldsen,Sonoi,Ball1,Ball2} as well as different reference models. The final inverted value is $0.01815\,\pm\,1 \times 10^{-4}$\,g\,cm$^{-3}$. Combining it with the radius determined from spectroscopic and astrometric data, we determine a model-independent mass interval of $1.17\,\pm\,0.11$\,M$_{\odot}$ for HD\,29399\footnote{While this mass interval is determined independently from any stellar model, it is however dependent on the accuracy of the radii values, hence on the accuracy of bolometric corrections, extinction laws, and so on.}.

\begin{figure*}[t]
\centering
        \adjincludegraphics[width=.9\textwidth, trim={0 0 0 0}, clip]{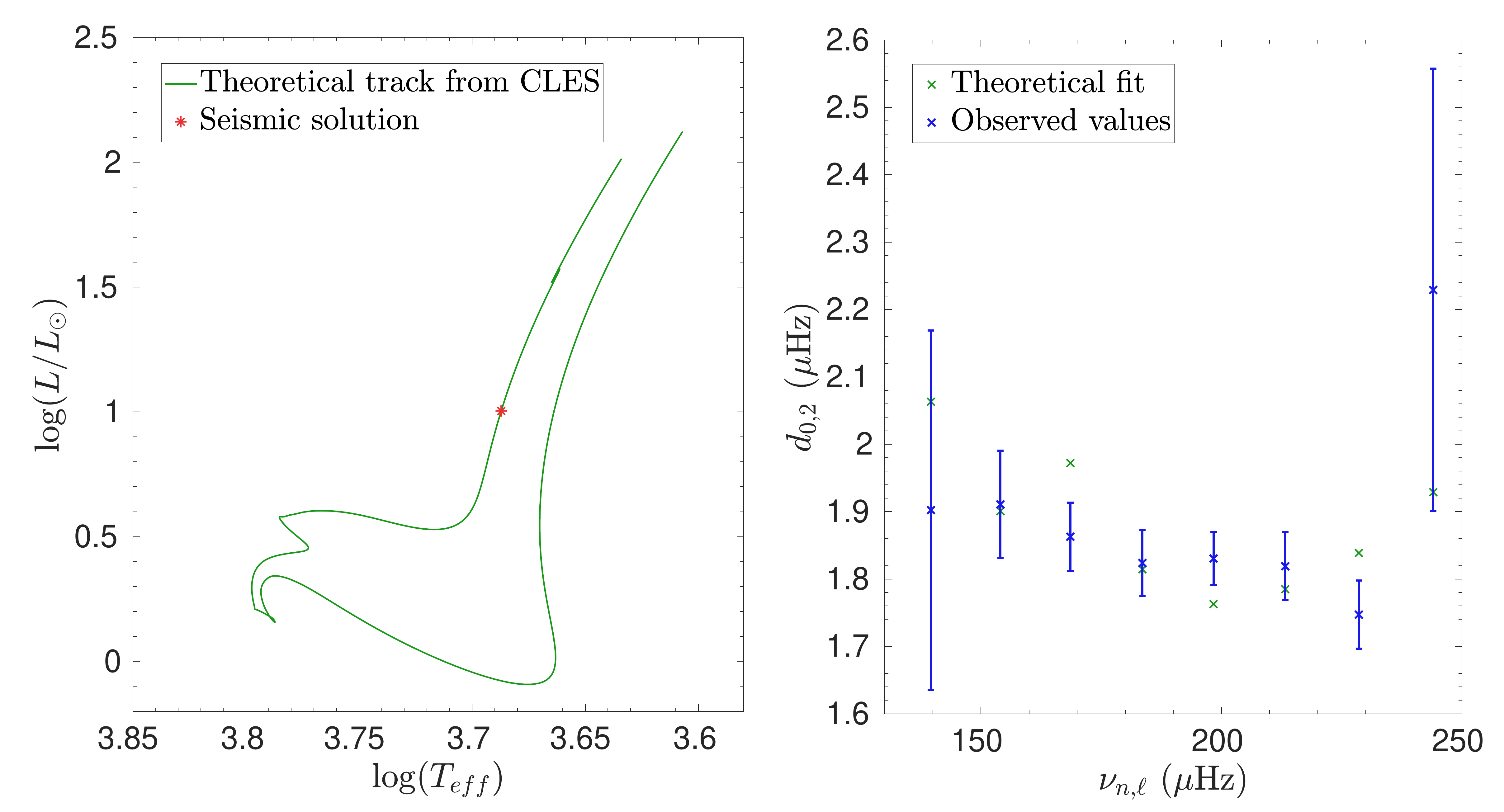}
        \caption{Left panel: HR diagram showing the CLES evolutionary track and the position of the optimal seismic solution. Right panel: Agreement of the $d_{0,2}$ for the optimal solution with the values determined from TESS data.}
        \label{FigTrackFit}
\end{figure*} 

The final modelling step is carried out using a Levenberg-Marquardt algorithm \citep[used e.g. in][]{BuldgenCygA, Farnir2019,Buldgen2019Kepler} with the following set of constraints: $\bar{\rho}_{\rm{Inv}}$, the inverted mean density, $L$, $T_{\rm{eff}}$, $\left[ \rm{Fe}/\rm{H}\right]$, and the individual small frequency separations, denoted $d_{0,2}=\nu_{n,0}-\nu_{n-1,2}$, using the mass, $M$, the age, $t$, the mixing-length parameter, $\alpha_{\rm{MLT}}$ , and the initial hydrogen, $X_{0}$, and metal content, $Z_{0}$, as free parameters. The stellar evolutionary models and their oscillation spectra are computed on the fly to avoid any interpolation issues in between and alongside the tracks. Departing from a solar-calibrated $\alpha_{\rm{MLT}}$ value implies additional degeneracies in the modelling. However, as we already have a good grasp on the expected mass of the star, we can ascertain that there will be a gain in accuracy, as departures from a solar-calibrated value are expected from the analysis of averaged $3D$ hydrodynamical simulations \citep{Trampedach2014, Magic, Sonoi2019}.


Using the small separations is justified as they allow mitigation of the surface effects, providing more robust estimates than those determined from individual frequency fitting. Moreover, the small separations have been shown by \citet{Montalban2010} to be very sensitive to the mass of the star as a function of its mean density, which serves our purpose of refining the stellar mass determination.

We illustrate the agreement of our final model with the constraints used in \autoref{FigTrackFit}, where we present on the left panel its evolutionary track showing its position on the lower RGB and on the right panel the agreement with the observed values of the individual $d_{0,2}$. The final parameters for the star are given in the second part of \autoref{tab:stellar_params}, where the reported uncertainties are determined from the analysis using the Levenberg-Marquardt minimisation. The precision of the stellar parameters is quite good, as the respective precisions of $d_{0,2}$, and the mean density are very high. However, it would be unrealistic to assume that this is the true precision of the determined stellar parameters. Varying the physical ingredients of the stellar evolution model may well lead to variations at a level comparable to the precision reported here. However, we note that this solution is well within the model-independent mass interval given by our inversion procedure. Therefore, below, we use the mass value from the second part of \autoref{tab:stellar_params} but consider the $1~\sigma$ uncertainty to be given by the combination of the inverted mean density and radius values from \textit{Gaia} parallaxes and spectroscopic parameters when determining the planetary parameters. 


\section{Orbital history}\label{SecOrbital}

In this section, we describe our study of the orbital evolution of the system. In addition to the non-rotating models of HD\,29399 computed with the CLES stellar evolution code, detailed models that include a comprehensive treatment of rotational effects and magnetic fields are computed with the Geneva stellar evolution code \citep{Eggenberger2008} from the PMS to the RGB phase. 



The stellar parameters determined through the seismic modelling described in the second part of \autoref{tab:stellar_params} are used as a starting point to compute these models. The assumption of shellular rotation \citep{zah92} is used and the internal transport of angular momentum (AM) is then solved simultaneously to the evolution of the star by accounting for meridional currents, transport by the shear instability, and transport by magnetic fields in the framework of the Tayler-Spruit dynamo \citep{spr02}. Consequently, the following equation for internal AM transport is solved:
\begin{equation}
  \rho \frac{{\rm d}}{{\rm d}t} \left( r^{2}\Omega \right)_{M_r} 
  =  \frac{1}{5r^{2}}\frac{\partial }{\partial r} \left(\rho r^{4}\Omega
  U(r)\right)
  + \frac{1}{r^{2}}\frac{\partial }{\partial r}\left(\rho (D_{\rm shear}+\nu_{\rm TS}) r^{4}
  \frac{\partial \Omega}{\partial r} \right) \, , 
\label{transmom}
\end{equation}
\noindent with $r$, $\rho(r)$, and $\Omega(r)$ being the radius, the mean density, and the mean angular velocity on an isobar, respectively. The radial dependence of the meridional circulation velocity in the radial direction is denoted $U(r)$, while $D_{\rm shear}$ corresponds to the diffusion coefficient for AM transport by the shear instability \citep[see][for more details]{egg10_rg}. The transport of AM by magnetic fields is taken into account through the viscosity $\nu_{\rm TS}$; this magnetic process is able to operate only when the shear parameter $q= -\frac{\partial \ln \Omega}{\partial \ln r}$ is larger than a minimum threshold given by $q_{\min}$ \citep[see][for mode details]{Eggenberger2019}. These models that account for rotational and magnetic effects are able to correctly reproduce the internal rotation of the Sun together with the observations of surface velocities of stars in open clusters \citep{Eggenberger2005,Eggenberger2019}. However, we note that the same models do not provide sufficient coupling to correctly reproduce the asteroseismic core rotation rates of red giants \citep{can14,den19}, which indicates that an unknown efficient additional AM transport process is needed for subgiant \citep{egg19} and red giant stars \citep{egg12_rg,egg17}.

\begin{figure}[t]
\centering
    \adjincludegraphics[width=1\columnwidth, trim={0 0 0 {.07\height}},clip]{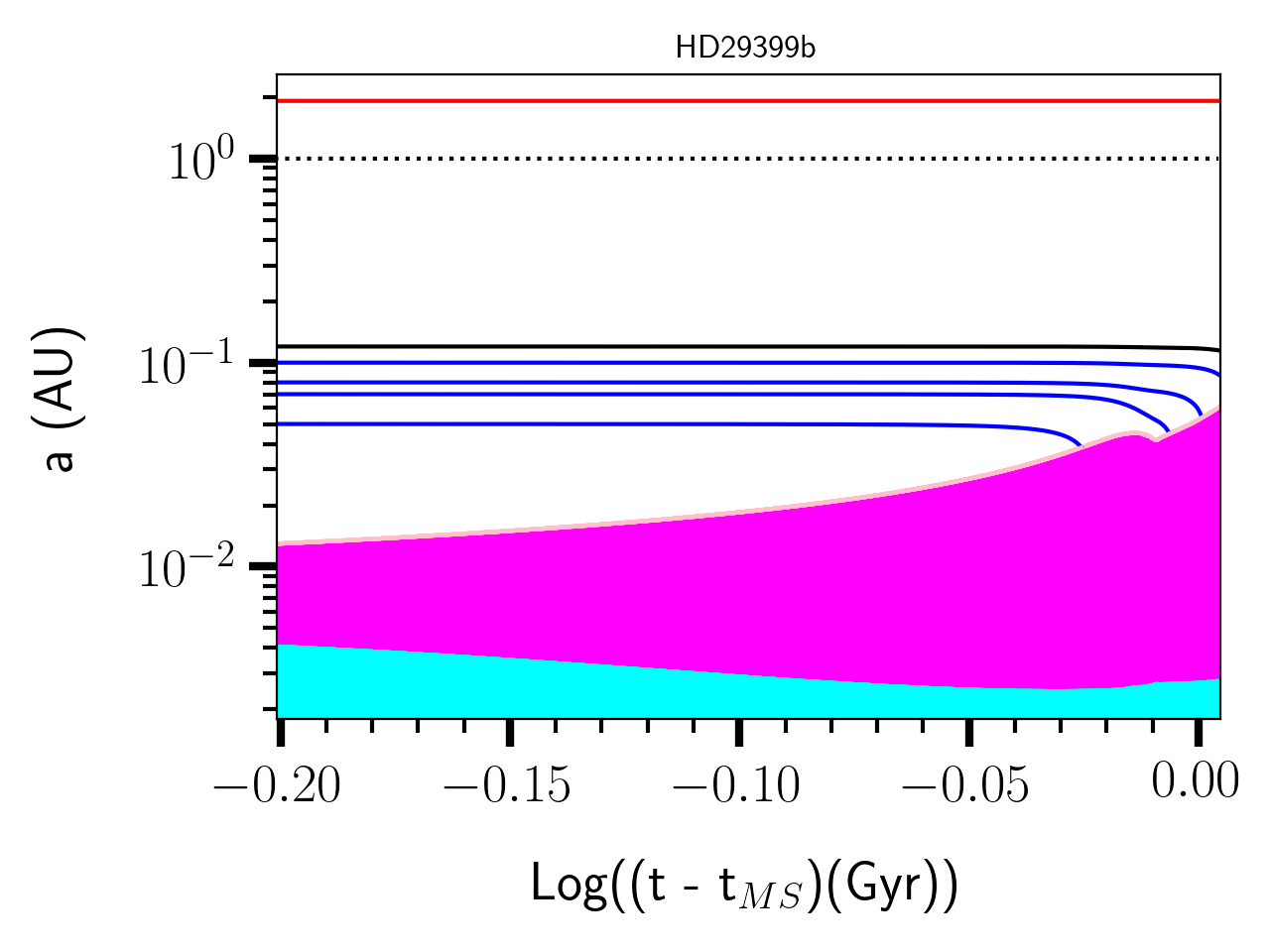}
    \caption{Evolution of the orbital distance after the MS (to the RGB branch). The solid red line represents the orbit of the planet. The solid black line shows the value for the orbital distance below which the planet will be engulfed. The magenta area represents the stellar convective envelope, while the cyan area shows the extension of the radiative interior.}
    \label{orbevo}
\end{figure}

To follow the evolution of a planetary system, these stellar models are coupled to our orbital evolution code by taking into account the exchange of angular momentum between the star and the orbit \citep{Privitera2016AII,Privitera2016III,Rao2018}. This enables us to test whether an eventual impact due to dynamical tides (mainly during the PMS) and/or equilibrium tides (at later evolutionary stages) could have significantly changed the orbit of the planet, leading to the current architecture of the system, or whether the system has retained the same architecture since its formation.

The physics included in the orbital evolution code is described in \cite{Rao2018}. We account for the planetary atmospheric evaporation occurring in Jeans escape or hydrodynamic escape regime conditions, depending on the properties of the system considered. For the computation of the emitted stellar X-ray luminosity, following the work by \citet{Tu2015}, we use the formula of \citet{Wright2011}, while for the EUV luminosity we refer to the work of \citet{SanzForcada2011}.\\

We started our study considering the planet at its current orbital distance from the host star at the beginning of the evolution, namely $\rm a_{in} = 1.910$ au, with a minimum mass $\rm M_{pl} = 1.59$ $\rm{M_J}$. The rotational history of the host star HD\,29399 being unknown, we considered three different initial surface angular velocities ($\rm 3.2$, $\rm 5$ and $\rm 18 \, \Omega_{\odot}$), covering the range for slow, medium, and fast rotators as deduced from surface rotation rates of solar-type stars observed in open clusters at different ages \citep{Eggenberger2019}. A disc lifetime of 6\,Myr is used for $\rm \Omega_{in} = 3.2, 5.0 ~ \Omega_{\odot}$ and 2\,Myr for $\rm \Omega_{in} = 18 ~ \Omega_{\odot}$. During the disc-locking phase, the surface angular velocity of the star is simply assumed to remain constant. After remaining constant during the disc-locking phase, the surface velocity rapidly increases due to the PMS contraction and reaches a peak at an age of about 25 Myr. Then the surface rotation rate exhibits a decrease along the evolution during the MS, indicating the braking of the stellar surface by magnetised winds. After the end of the MS, the surface rotation decreases rapidly due to the expansion of the envelope during the subgiant and the red giant phases. 


Using the initial setup described above, we did not find any appreciable change in the orbit of the planet along the evolution. Given the mass of the planet and that of the star, the planet is found to be at too great an  initial distance for its motion to be significantly affected  by tides. Even during the RGB phase, the limited increase in the stellar radius (a value of only about $\rm 12.9 \, R_{\odot}$ is found for HD\,29399, corresponding to $\rm 0.0586 $ au) does not give rise to efficient equilibrium tides. These results are in perfect agreement with previous results reported in \cite{Privitera2016a}.

For the sake of completeness, we computed the orbital evolution of the planet when the host star climbs the RGB, exploring a range of lower initial orbital distances to determine the maximal value below which the planet would be engulfed, denoted here $\rm{a_{Max}}$. As shown in \autoref{orbevo}, we find a value of $\rm{a_{Max}} \approx$\,0.12\,au. In the case of HD\,29399, this is more than 15 times lower than the observed orbital distance.


\section{Conclusion}\label{SecConc}

In this paper, we illustrate the advantages of synergy between both exoplanetary studies and seismic characterisation of stars with the detailed characterisation of a long-period giant planet orbiting the evolved star HD\,29399 detected within the 14-year CASCADES survey (Ottoni et al.  2021, submitted), and observed by TESS during 11 continuous sectors of 27 days. The detailed analysis allows us to provide very precise stellar parameters, far beyond what is achievable from stellar evolutionary tracks alone, paving the way for an analysis of the orbital evolution of the planetary system.

The newly discovered giant planet is a 1.57\,M$_J$ companion at an orbital distance of  1.913\,au, with a period of 892.7\,days. The radial-velocity time-series also present a long-term linear trend that may reveal a substellar companion. We consulted the corresponding ASAS-3 photometry time-series \citep{Pojmanski2002} to address the announcement of a false-positive by \citet{Wittenmyer2017a} and demonstrate that the photometric time-series must be considered carefully, and cannot not rule out the planet hypothesis. We also checked for any correlation with chromospheric activity and spectroscopic line-profile variations, and find no significant signals.

In addition to the exquisite coverage of radial velocities, HD\,29399 has also been followed for almost a year by TESS, allowing for the precise determination of its pulsational properties. This allowed us to determine individual radial and quadrupolar oscillation modes with high precision and use them for a detailed modelling of the host star. This allowed us to determine the mass of HD\,29399 to be $1.17\pm0.10$\,M$_{\odot}$, the radius to be $4.47\pm 0.02\, \rm{R}_{\odot}$, and the age of the system to be around $6.2 \pm 0.5 \,\rm{Gyr}$.

The determination of these detailed properties enabled us to study the orbital evolution of the system using non-standard stellar evolution models computed with the Geneva stellar evolution code. Our study shows that neither dynamical nor equilibrium tides have been able to affect the orbital evolution of the planet, whatever the initial rotational velocity considered for the host star. Our results are in agreement with those of \citet{Privitera2016AII}, showing that a Jupiter-mass planet orbiting at such long periods should remain unaffected. We carried out our computations up to the  tip of the RGB and predict no engulfment in the future evolution of the system. 

Overall, our study demonstrates a perfect example of a multidisciplinary, multi-instrument approach, where a long-duration ground-based survey combined with high-cadence space-based observations   has led to the  discovery of a long-period planet and precise asteroseismic characterisation of its host star. We combined precise stellar and planetary characterisations in our study of the orbital evolution of the system. Such multidisciplinary approaches are crucial to understanding the statistical properties of planetary systems, their formation history, and the evolution of their properties as a result of interactions with their host star. 

\begin{acknowledgements}
We thank the referee for the useful comments that helped to improve the quality of the manuscript. This work has been carried out within the framework of the National Centre for Competence in Research PlanetS supported by the Swiss National Science Foundation. The authors acknowledge the financial support of the SNSF. \newline
This publication makes use of the The Data \& Analysis Center for Exoplanets (DACE), which is a facility based at the University of Geneva (CH) dedicated to extrasolar planet data visualisation, exchange and analysis. DACE is a platform of the Swiss National Centre of Competence in Research (NCCR) PlanetS, federating the Swiss expertise in Exoplanet research. The DACE platform is available at https://dace.unige.ch. \newline
We thank all observers at La Silla Observatory from the past fourteen years for their quality work. We acknowledge financial support from the Swiss National Science Foundation (SNSF). This work has, in part, been carried out within the framework of the National Centre for Competence in Research PlanetS supported by SNSF.\\
C.P. acknowledges funding from the Swiss National Science Foundation (project Interacting Stars, number 200020-172505). G.B. acknowledges funding from the SNF AMBIZIONE grant No 185805 (Seismic inversions and modelling of transport processes in stars). P.E. has received funding from the European Research Council (ERC) under the European Union's Horizon 2020 research and innovation programme (grant agreement No 833925, project STAREX). This article used an adapted version of InversionKit, a software developed within the HELAS and SPACEINN networks, funded by the European Commission's Sixth and Seventh Framework Programmes. AM acknowledges support from the ERC Consolidator Grant funding scheme (project ASTEROCHRONOMETRY, https://www.asterochronometry.eu, G.A. n. 772293).
\end{acknowledgements}

\bibliography{biblioarticleTess}

\appendix
\section{Radial-velocity data}
\begin{table}[ht]
\centering
\caption{Radial-velocity measurements and uncertainties for HD\,29399 obtained with the CORALIE and HARPS spectrographs.}
\begin{threeparttable}
    \begin{tabular}{llll}        
    \hline
    \hline
    JD-2\,400\,000 & RV & e\_RV & Instrument\\
    & $[m\,s^{-1}]$ & $[m\,s^{-1}]$ & \\
54035.785459 & 31627.21 & 3.25 & CORALIE98\\
54371.782019 & 31694.66 & 5.69 & HARPS\\
54472.673890 & 31643.86 & 5.06 & CORALIE07\\
54689.929529 & 31619.75 & 2.72 & CORALIE07\\
54764.767582 & 31637.14 & 4.96 & CORALIE07\\
54772.828843 & 31650.47 & 3.67 & CORALIE07\\
54777.770615 & 31635.50 & 3.91 & CORALIE07\\
54817.667858 & 31644.19 & 5.27 & CORALIE07\\
54830.700228 & 31646.48 & 2.98 & CORALIE07\\
54892.579440 & 31677.34 & 5.74 & HARPS\\
54894.550719 & 31686.96 & 5.69 & HARPS\\
55176.740700 & 31676.88 & 2.48 & CORALIE07\\
55303.503222 & 31696.39 & 5.20 & CORALIE07\\
55446.907968 & 31642.22 & 2.45 & CORALIE07\\
55619.560825 & 31625.53 & 2.54 & CORALIE07\\
55858.875008 & 31680.45 & 2.86 & CORALIE07\\
55929.675679 & 31692.51 & 2.31 & CORALIE07\\
56952.825823 & 31709.20 & 8.81 & CORALIE07\\
\hline
57346.699387 & 31648.69 & 2.29 & CORALIE14\\
57763.623634 & 31714.20 & 2.90 & CORALIE14\\
57829.493845 & 31725.52 & 2.64 & CORALIE14\\
57971.892631 & 31697.45 & 2.73 & CORALIE14\\
58031.714400 & 31675.10 & 2.96 & CORALIE14\\
58094.561422 & 31674.10 & 2.38 & CORALIE14\\
58150.558636 & 31645.80 & 8.71 & HARPS\\
58172.514760 & 31657.57 & 2.59 & CORALIE14\\
58178.551656 & 31673.40 & 2.75 & CORALIE14\\
58323.934316 & 31652.97 & 2.58 & CORALIE14\\
58422.854271 & 31673.54 & 3.63 & CORALIE14\\
58482.776199 & 31691.53 & 2.73 & CORALIE14\\
58775.738473 & 31718.29 & 3.07 & CORALIE14\\
58846.731357 & 31697.70 & 2.68 & CORALIE14\\
\hline
    \end{tabular}
\begin{tablenotes}
    \footnotesize
    \item Note that small radial-velocity offsets between the different versions of each instruments have to be considered. In this case, the offset between COR98 and COR07 has been fixed at 0\,m\,s$^{-1}$, and the offset between COR14 and COR07 has been considered as a free parameters in the model (see \autoref{tab:orbit_params}).
\end{tablenotes}
\end{threeparttable}
\label{tab:timeseries_hd29399}
\end{table}


\newpage

\section{MCMC - corner plot distributions of fitted parameters}\label{apdx:corner}
 
\begin{figure*}[h]
        \centering
        \adjincludegraphics[width=\textwidth, trim={0 0 0 0},clip]{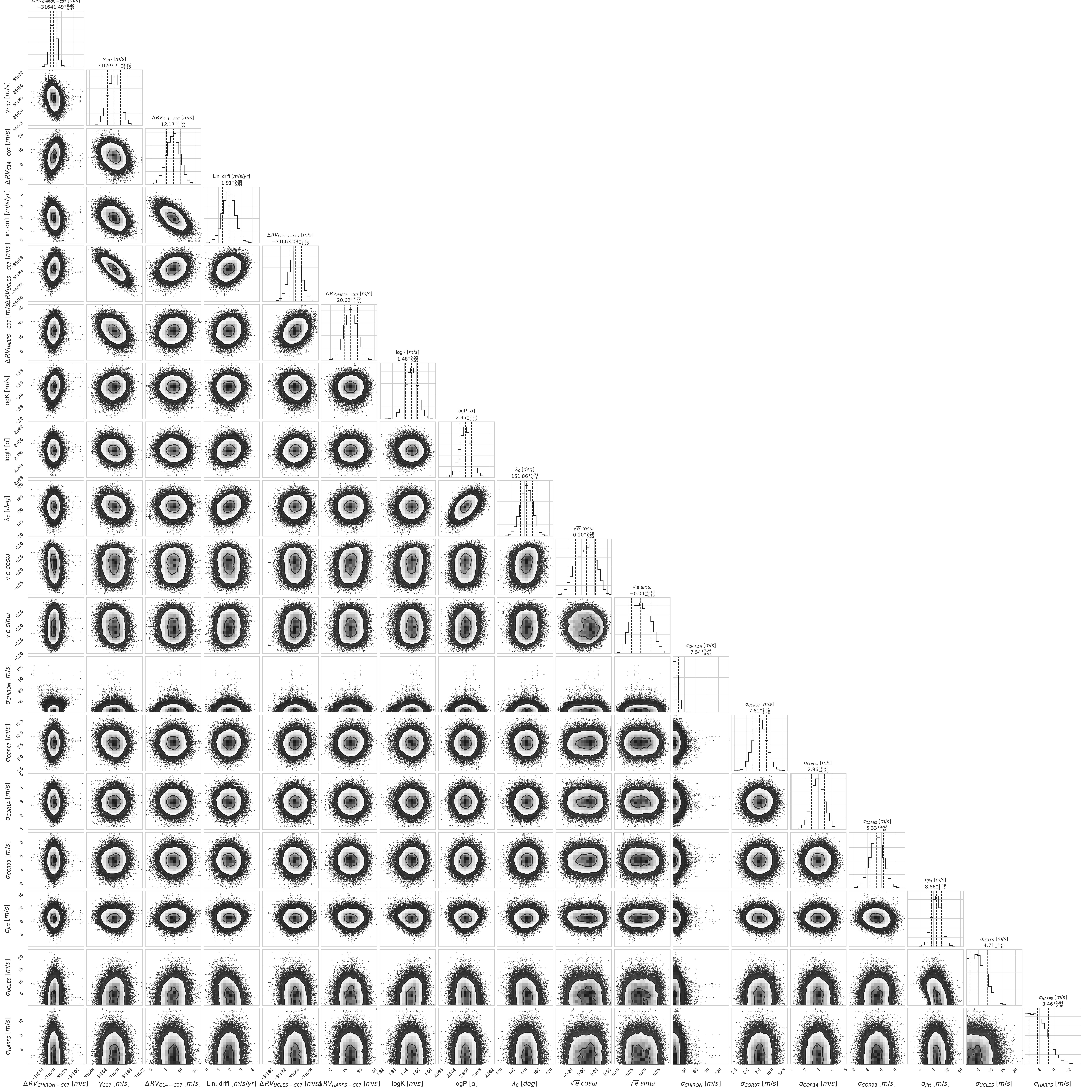}\\
        \caption{Posterior distributions of the fitted parameters of HD29399. Each panel contains the two-dimensional histograms of the 1\,200\,000 samples (after removal of the burn-in, the first 25\% of the chains). Contours are drawn to improve the visualisation of the 1$\sigma$ and 2$\sigma$ confidence interval levels. The upper panels of the corner plot show the probability density distributions of each orbital parameter of the final MCMC sample. The vertical dashed lines mark the 16th, 50th and the 84th percentiles of the overall MCMC samples, delimiting the 1$\sigma$ confidence interval.  }
        \label{fig:hd29399_corner}
\end{figure*}

\end{document}